\documentclass[prodmode,hillsideplop]{acmlarge}
\usepackage{flafter}
\usepackage{hyperref}
\usepackage[hyphenbreaks]{breakurl}

\usepackage[ruled]{algorithm2e}
\SetAlFnt{\algofont}
\SetAlCapFnt{\algofont}
\SetAlCapNameFnt{\algofont}
\SetAlCapHSkip{0pt}
\IncMargin{-\parindent}

\title{Applying Software Patterns to Address Interoperability 
in Blockchain-based Healthcare Apps}
\author{Peng Zhang \affil{Vanderbilt University, Nashville, TN}\ \\
Jules White \affil{Vanderbilt University, Nashville, TN}\ \\
Douglas C. Schmidt \affil{Vanderbilt University, Nashville, TN}\ \\
Gunther Lenz \affil{Varian Medical Systems, Palo Alto, CA}\ \\
}

\begin{abstract}
Since the inception of the Bitcoin technology, its underlying data structure--the blockchain--has garnered much attention due to properties such as decentralization, transparency, and immutability. These properties make blockchains suitable for apps that require disintermediation through trustless exchange, consistent and incorruptible transaction records, and operational models beyond cryptocurrency.  In particular, blockchain and its smart contract capabilities have the potential to address healthcare interoperability issues, such as enabling effective interactions between users and medical applications, delivering patient data securely to a variety of organizations and devices, and improving the overall efficiency of medical practice workflow. Despite the interest in using blockchain for healthcare interoperability, however, little information is available on the concrete architectural styles and patterns for applying blockchain to healthcare apps.  This paper provides an initial step in filling this gap by showing: (1) the features and implementation challenges in healthcare interoperability, (2) an end-to-end case study of a blockchain-based healthcare app we are developing, and (3) how applying foundational software patterns can help address common interoperability challenges faced by blockchain-based healthcare apps.
\end{abstract}

\category{H.5.2}{Software Patterns}{Blockchain-based healthcare apps}
\copyr{}

\begin{document}

% \begin{bottomstuff}
% This work is supported by the Widget Corporation Grant \#312-001.\\
% Author's address: D.\ Pineo, Kingsbury Hall, 33 Academic Way, Durham,
% N.H. 03824; email: dspineo@comcast.net; Colin Ware, Jere A.\ Chase
% Ocean Engineering Lab, 24 Colovos Road, Durham, NH 03824; email: cware@ccom.unh.edu\\

% Permission to make digital or hard copies of all or part of this work for personal or classroom use is granted without fee provided that copies are not made or distributed for profit or commercial advantage and that copies bear this notice and the full citation on the first page. To copy otherwise, to republish, to post on servers or to redistribute to lists, requires prior specific permission. A preliminary version of this paper was presented in a writers' workshop at the 20th Conference on Pattern Languages of Programs (PLoP).
% \end{bottomstuff}

\maketitle

\section{Introduction}
Over the past several years blockchain technology has attracted interest from computer scientists and domain experts in various industries, including finance, real estate, healthcare, and transactive energy. This interest initially stemmed from the popularity of Bitcoin~\cite{nakamoto2012bitcoin} and the Bitcoin platform, which is a cryptographic currency framework that was the first application of blockchain.  Blockchain possesses certain properties, such as decentralization, transparency, and immutability, that have allowed Bitcoin to become a viable platform for "trustless" transactions, which can occur directly between any parties without the intervention of a centralized intermediary. 

Another blockchain platform, Ethereum, extended the capabilities of the Bitcoin blockchain by adding support for "smart contracts." Smart contracts are computer programs that directly control exchanges or redistributions of digital assets between two or more parties according to certain rules or agreements established between involved parties.  Ethereum's programmable smart contracts enable the development of decentralized apps (DApps)~\cite{johnston2014general} , which are autonomously operated services cryptographically stored on the blockchain that enable direct interaction between end users and providers. 
 
This paper focuses on a previously unexplored topic related to blockchain, namely, the application of software patterns to modularize and facilitate extensibility of blockchain-based apps that focus on addressing the interoperability challenges in healthcare.  In the healthcare context, interoperability is defined as the ability for different information technology systems and software apps to communicate, exchange data, and use the information that has been exchanged. The high (and growing) cost of healthcare in many countries motivates our focus on applying blockchain technologies to help bridge the gap in communication and information exchange~\cite{desalvo2015connecting}.

The remainder of this paper is organized as follows: Section 2 gives an overview of blockchain and outlines how blockchain-based apps can help address key challenges in healthcare interoperability; Section 3 provides an end-to-end case study of a blockchain-based healthcare app we are developing and the implementation challenges we encountered when extending the app; Section 4 describes foundational software patterns that can be applied to address interoperability requirements in the blockchain app covered in Section 3 and discusses key concerns when realizing these patterns in healthcare-focused blockchain apps; Section 5 compares our work with existing work on potential pros and cons of a health blockchain; and Section 6 presents concluding remarks and summarizes our future work on applying blockchain technologies in the healthcare domain.

\section{Overview Blockchain and Its Role in Healthcare Apps}
This section gives an overview of blockchain and the open-source Ethereum blockchain that provides additional support for smart contracts.  It then outlines key challenges in healthcare with respect to interoperability and how the properties of blockchain-based apps can help address these challenges

\subsection{Overview of Blockchain}
\begin{figure}[bp]
\centering
\includegraphics[width=0.5\textwidth]{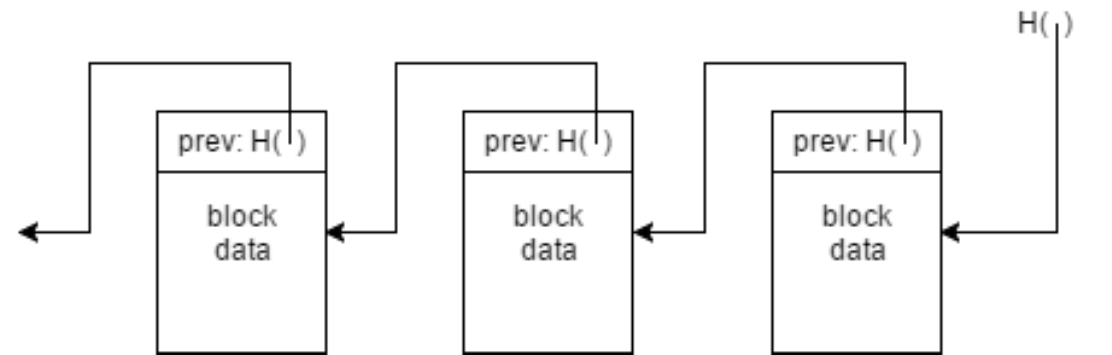}
\caption{Blockchain Structure: a Continuously Growing List of Ordered and Validated Transactions}
\label{blockchain}
\end{figure}

A blockchain is a decentralized computing architecture that maintains a growing list of ordered transactions grouped into blocks that are continually reconciled to keep information up-to-date, as shown in Figure ~\ref{blockchain}.  Only one block can be added to the blockchain at a time and each block is mathematically verified (using cryptography) to ensure it follows in sequence from the previous block to maintain consensus across the entire decentralized network.  The verification process is also called "mining" or Proof of Work (PoW)~\cite{nakamoto2012bitcoin}, which allows network nodes (also called "miners") to compete to be the first to have their block be the next one added to the blockchain by solving a computationally expensive puzzle.  The winner then announces the solution to the entire network to gain some mining rewards in cryptocurrency.  This mechanism combines game theory, cryptography, and incentive engineering to ensure that the network reaches consensus regarding each block in the blockchain and that no tampering occurs with the transaction history.  All transaction records are kept in the blockchain and are shared with all network nodes, thereby ensuring properties of transparency, incorruptibility, and robustness (since there is no single point of failure).

In the Bitcoin application, a blockchain serves as a public ledger for all transactions of cryptocurrency in bitcoins to promote trustless finance between individual users, securing all their interactions with cryptography.  The Bitcoin blockchain has limitations, however, when supporting different types of applications involving contracts, equity, or other information, such as crowdfunding, identity management, and democratic voting registry~\cite{buterin2014ethereum}.  To address the needs for a more flexible framework, Ethereum was created as an alternative blockchain, giving users a general, trustless platform that can run smart contracts, which are computer protocols that enable different types of decentralized applications beyond cryptocurrencies.  

The Ethereum public blockchain is essentially a distributed state transition system, where state is made up of accounts and state transitions are direct transfers of value and information between accounts. Two types of accounts exist in Ethereum: (1) \textit{externally owned accounts} (EOAs), which are controlled via private keys and only store Ethereum's native value-token "ether" and (2) \textit{smart contract accounts} (SCAs) that are associated with contract code and can be triggered by transactions or function calls from other contracts~\cite{buterin2014ethereum}.

To protect the blockchain from malicious attacks and abuse (such as distributed denial of service attacks in the network or hostile infinite loops in the contract code), Ethereum also enforces a payment protocol, whereby a fee is charged for memory storage and each computational step that is executed in a contract or transaction.  These fees are collected by miners who verify, execute, propagate transactions, and then group them into blocks. Just like in the Bitcoin network, the mining rewards provide an economic incentive for users to dedicate powerful hardware and electricity to the public Ethereum network.

\subsection{Addressing Healthcare Interoperability via Blockchain-based Apps}

% Conventional EHR systems provide a partial solution that facilitates patient care management within a health organization. Most existing EHR systems, however, are incompatible with each other, hindering interactions between diverse care providers across the nation.  For example, transferring data from one system to another causes inefficiencies in the medical practice workflow that increases costs and degrades the quality of patient care.
	
% Basic technical requirements for achieving interoperability, defined by the Office of the National Coordination for Health Information Technology (ONC), include identifiability and authentication of all participants, ubiquitous and secure infrastructure to store and exchange data, authorization and access control of data sources, and the ability to handle data sources of various structures~\cite{onc2014vision}.  Blockchain technologies are emerging as promising and cost-effective means to meet some of these requirements due to their inherent design properties, such as secure cryptography and a resilient peer-to-peer network.  Likewise, blockchain-based apps can benefit the healthcare domain via their properties of asset sharing, audit trails of data access, and pseudonymity for user information protection, which are essential for solving interoperability issues in healthcare. 

Many research and engineering ideas have been proposed to apply blockchain to healthcare and implementation attempts underway~\cite{ekblaw2016case,peterson2016blockchain,porru2017blockchain,bartoletti2017empirical}, but few published studies have addressed the software design considerations needed to implement blockchain-based healthcare apps effectively.  While it is crucial to understand the fundamental properties that blockchain possesses, it is also important to apply solid software engineering practices when programming the blockchain to minimize software maintenance effort via code modularity and extensibility, especially in the fast-growing and highly demanded healthcare domain.  The remainder of this section summarizes key interoperability challenges in healthcare and how blockchain technologies can provide assistance.

\subsubsection{Challenge 1: Maintaining Evolvability While Minimizing Integration Complexity}
 
Many applications are written with the assumption that data is easy to change. In a blockchain application, however, data is immutable and difficult to modify in mass. A critical design consideration when building blockchain applications for healthcare is ensuring that the data and contracts written into the blockchain are designed in a way to facilitate evolution where needed. Although evolution must be facilitated, healthcare data must often be accessible from a variety of deployed systems that cannot easily be changed over time. Therefore, the evolvability should be designed in a way that minimizes the impact of evolution on the clients that interact with data in the blockchain. Section 4.2.1 shows how the Abstract Factory pattern can be applied in Ethereum contracts to facilitate evolution while minimizing the impact on dependent healthcare application clients.
 
\subsubsection{Challenge 2: Minimizing Data Storage Requirements}
 
Healthcare applications can serve thousands or millions of participants, which can potentially place an enormous burden when all of this data is stored in the blockchain -- particularly if data normalization and denormalization techniques are not carefully thought through. An important design consideration is maximizing data sharing while ensuring that sufficient flexibility exists to manage individual health concerns. As we describe in Section 4.2.2, the Flyweight pattern can be applied to help ensure that common intrinsic data is shared across Ethereum contracts while still allowing extrinsic data to vary across the contracts specific to individuals.
 
\subsubsection{Challenge 3: Balancing Integration Ease with Security Concerns}
 Basic technical requirements for achieving interoperability, defined by the Office of the National Coordination for Health Information Technology (ONC), include identifiability and authentication of all participants, ubiquitous and secure infrastructure to store and exchange data, authorization and access control of data sources, and the ability to handle data sources of various structures~\cite{onc2014vision}.  Blockchain technologies are emerging as promising and cost-effective means to meet some of these requirements due to their inherent design properties, such as secure cryptography and a resilient peer-to-peer network.  Likewise, blockchain-based apps can benefit the healthcare domain via their properties of asset sharing, audit trails of data access, and pseudonymity for user information protection, which are essential for solving interoperability issues in healthcare. Although there are substantial potential benefits to interoperability and availability of storing patient data in the blockchain, it carries significant risks -- even when encryption is applied. In Section 4.2.3, we discuss how the Proxy pattern can be applied to aid in facilitating interoperability through the blockchain while keeping sensitive patient data from being directly encoded in the blockchain.  
 
\subsubsection{Challenge 4: Tracking Relevant Health Changes Across Large Patient Populations}
 
Communication gaps and information sharing challenges are a serious impediment to healthcare innovation and the quality of patient care.  Providers, hospitals, insurance companies, and departments within health organizations experience disconnectedness caused by delayed or lack of information flow.  It is common for patients to be cared for by various sources like private clinics, regional urgent care centers, and enterprise hospitals. As we discuss in Section 4.2.4, a provider may have hundreds or more patients that have associated contracts that they need to track. Determining how to scalably detect when changes of relevance in an individual's contract have occurred is not easy but can be addressed by implementing the Publisher Subscriber pattern in Ethereum contracts. 

\section{Case Study of the DApp for Smart Health (DASH)}
This section presents the structure and functionality of the DASH app we are developing to explore the efficacy of applying blockchain technology to the healthcare domain.  It also discusses the challenges associated with developing blockchain-based apps to improve healthcare interoperability.

\subsection{Structure and Functionality of DASH}
\begin{figure}[bp]
\centering
\includegraphics[width=0.5\textwidth]{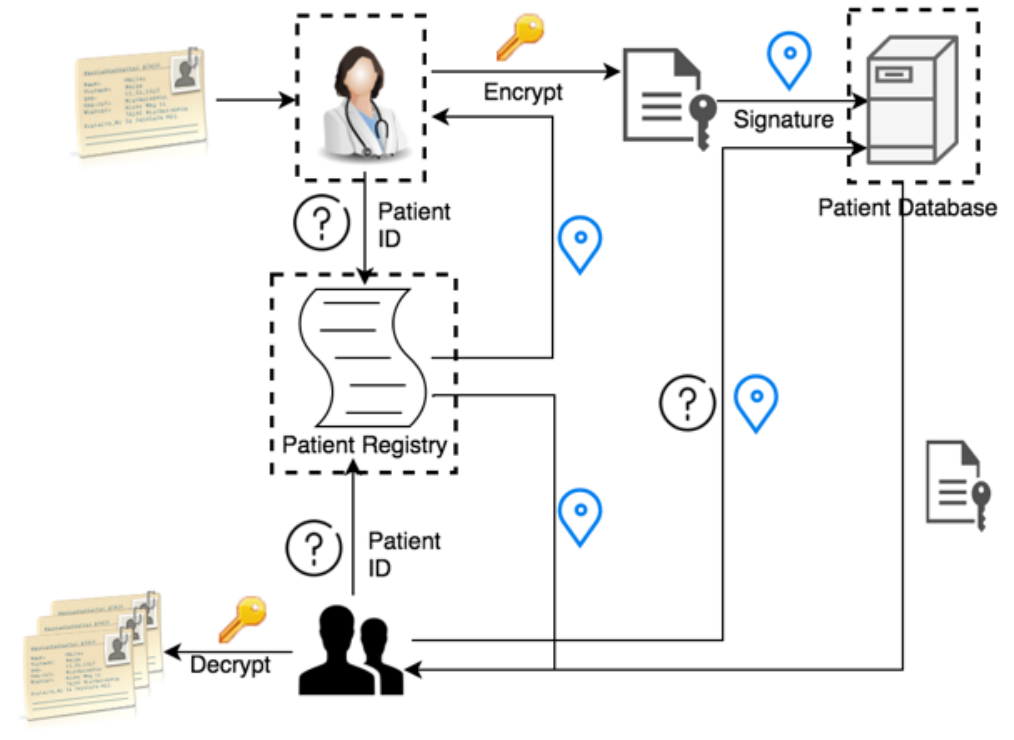}
\caption{Structure and Workflow of DASH}
\label{dash}
\end{figure}

DASH provides a web-based portal for patients to access and update their medical records, as well as submit prescription requests.  Likewise, the app allows providers to review patient data and fulfill prescription requests based on permissions given by patients.  Figure~\ref{dash} gives an overview of the structure of DASH.

DASH is implemented on an Ethereum test blockchain, with a SMART (Substitutable Medical Apps, Reusable Technology)~\cite{mandel2016smart} on FHIR (Fast Healthcare Interoperability Resources)~\cite{bender2013hl7} schema as the standard data format for stored patient data.  Using the smart contract support in Ethereum, a \textit{Patient Registry} contract is created to store a mapping between unique patient identifiers and their associated \textit{Patient Account} contract addresses.  Each \textit{Patient Account} contract also contains a list of health providers that are granted read/write access to the patient's medical records.  At its initial state, DASH can provide basic access services to the two types of users: patients and providers. 

\subsection{Key Implementation Challenges of DASH}
Although our prototype of DASH was effective for its initial purposes, the following challenges arose when we attempted to extend it to support new types of users from other departments within the same organization:
\begin{enumerate}
\item \textbf{Tightly coupled designs.} Our initial design created tight coupling between a lot of variable components, resulting in rewriting of many contracts that propagated to changes in client code.  For instance, metadata can be stored in the smart contract internal states as member variables or event logs, but when better design choices are available, new contracts must be instantiated, leaving existing contracts obsolete due to blockchain's immutability property.  Unfortunately, the tight coupling between (1) storage and access of storage and (2) entities and entity creation mechanism caused significant overhead in terms of implementation effort as well as computational and storage costs.  
\item \textbf{Duplicated resources.} Despite the diverse functioning entities in healthcare, much commonality may be shared across multiple entities.  For example, patients assigned to different care management teams may have the same insurance plans, which could be duplicated many times if their care teams belong to different departments, consuming more than necessary storage to maintain.  Duplicated resources are also difficult and costly to manage because when an update occurs, all the corresponding copies have to be updated correctly and timely to avoid confusion.  In the likely event of failing to update certain copies, all copies will have to be reevaluated or recollected to ensure data integrity. 
\item \textbf{Lack of scalability.} Another challenge in blockchain-based healthcare apps stems from the ease of broadcasting events to separate functioning groups, such as patients, providers, and billing departments.  For instance, to update providers on a particular patient's activities, a back-end server must perform computationally-intense tasks to constantly watch for events associated with that patient's \textbf{Patient Account} contract.  This approach, unfortunately, does not scale when a large number of parties need to receive notifications regarding activities of other parties with which they interact.
\end{enumerate}

To resolve these implementation challenges while still providing solutions to address the interoperability challenges in healthcare as aforementioned, we redesigned the DASH app with better software engineering practice by applying foundational software patterns as we discuss in Section 4. 

\section{Applying Foundational Software Patterns to Blockchain-based Health Apps}
This section describes how foundational software patterns~\cite{gamma1995design,buschmann2007pattern} can be applied in blockchain-based health apps (such as DASH) to address the interoperability challenges described in Section 2 while providing solutions to the implementation concerns highlighted in Section 3.  
% This section addresses how foundational software patterns~\cite{gamma1995design,buschmann2007pattern} can be applied in a blockchain-based health app to address important concerns from a healthcare perspective when reifying these patterns.

\subsection{Overview of Solidity}
Ethereum smart contracts are built on a Turing-complete programming language, called Solidity~\cite{solidity2017ethereum}. This contract language has allowed the Ethereum blockchain to become a platform for creating decentralized applications (DApps), thereby providing possible solutions to healthcare interoperability challenges.   

The Solidity language has an object-oriented flavor and is intended primarily for writing contracts in Ethereum.  A "class" in Solidity is realized through a "contract", which is a prototype of an object that lives on the blockchain.  Just like an object-oriented class can be instantiated into a concrete object at runtime, a contract may be instantiated into a concrete SCA by a transaction or a function call from another contract.  At instantiation, it is given a uniquely identifying address similar to a reference or pointer in C/C++-like languages, with which it can then be called.  Contracts also contain persistent state variables that can be used as data storage.  Although one contract can be instantiated into many SCAs, it should be treated as a singleton to avoid storage overhead.  After a contract is created, its associated SCA address is typically stored at some place (e.g., a configuration file or a database) and used by the application to access its internal states and functions.  

Solidity supports multiple inheritance including polymorphism.  When a contract inherits from one or more other contracts, a single contract is created by copying all the base contracts code into the created contract instance.  Abstract contracts in Solidity allow declaration headers to be defined without concrete implementations.  They cannot be compiled into a SCA but can be used as base contracts.  Due to their similarity to C/C++-like classes, many foundational software patterns can be applied to smart contracts to address various design challenges, as described next.

\subsection{Applying Software Patterns to Improve DASH Design}
\begin{figure}[bp]
\centering
\includegraphics[width=0.6\textwidth]{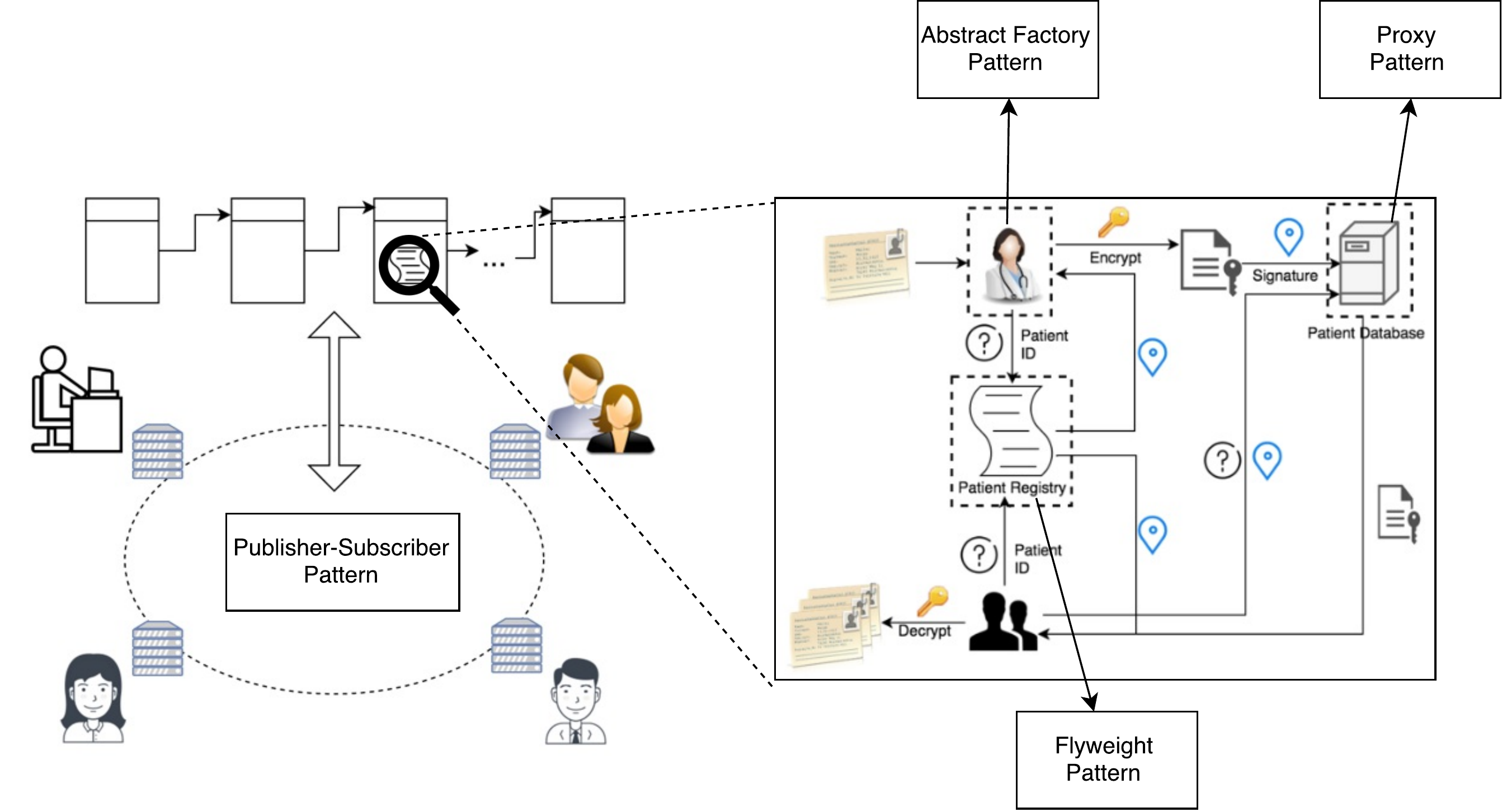}
\caption{Applying Software Patterns to Improve DASH Design and Extensibility}
\label{patterns}
\end{figure}

The remainder of this section focuses on four software patterns--Abstract Factory, Flyweight, Proxy, and Publisher-Subscriber--that we applied to DASH to address the healthcare interoperability challenges in Section 2, but they are not the only patterns relevant in this domain.  Figure~\ref{patterns} shows how these patterns can interact in the DASH app's ecosystem.  For instance, \textit{Abstract Factory} can assist with user account creations based on user types; whereas \textit{Flyweight} ensures accounts on the blockchain are unique and maximizes commonality sharing, and \textit{Publisher-Subscriber} can be used to notify collaborating users when events of interest occur on the blockchain. In addition, to abstract away storage implementation detail, the Proxy pattern can be applied to allow seamless interactions between separate components in the system while still supporting variations in data storage options.

\subsubsection{Applying Abstract Factory to Address Interoperability Challenge 1}\mbox{}\\

\textbf{Motivation.}  Smart contracts allow code to be executed on the blockchain by the EVM.  The incorruptibility property of blockchain ensures that interfaces (e.g. methods and properties) of already instantiated contracts cannot be modified or upgraded.  Each new version of a smart contract has to be created as a new contract object on the blockchain and distributed among all the network nodes to be executed on demand.  Therefore, it is important to design a contract class to modularize code and minimize changes to its interface over time.

For example, if a blockchain is used to enable some interactions between different departments in a hierarchical health organization, without a well thought out design, a lot of decisions have to be made by clients based on the specific department or sub-department type (a large number of if-else statements) encountered.  As new departments are introduced, the decision-making code will likely have to change more than once, and each obsolete version of contract discarded.

The Abstract Factory pattern can facilitate this scenario because its "factory" object (the factory itself is a contract instance) is then responsible for providing creation services of concrete departmental contracts for the entire health organization.  Each factory object can create contracts for a group of departments or subdivisions that are related or always interact, and it is easy to instantiate another factory object when new interactions take place.  As a more concrete example, the billing department within a hospital organization may have smaller subdivisions that assist different sets of patients depending on what insurance they carry.  When the hospital decides to accept insurance plans from a new provider, the interaction between the hospital subdivision and patients with the new provider can simply be created by a factory instead of changing the existing code to include this new interaction.    

\begin{figure}[bp]
\centering
\includegraphics[width=0.6\textwidth]{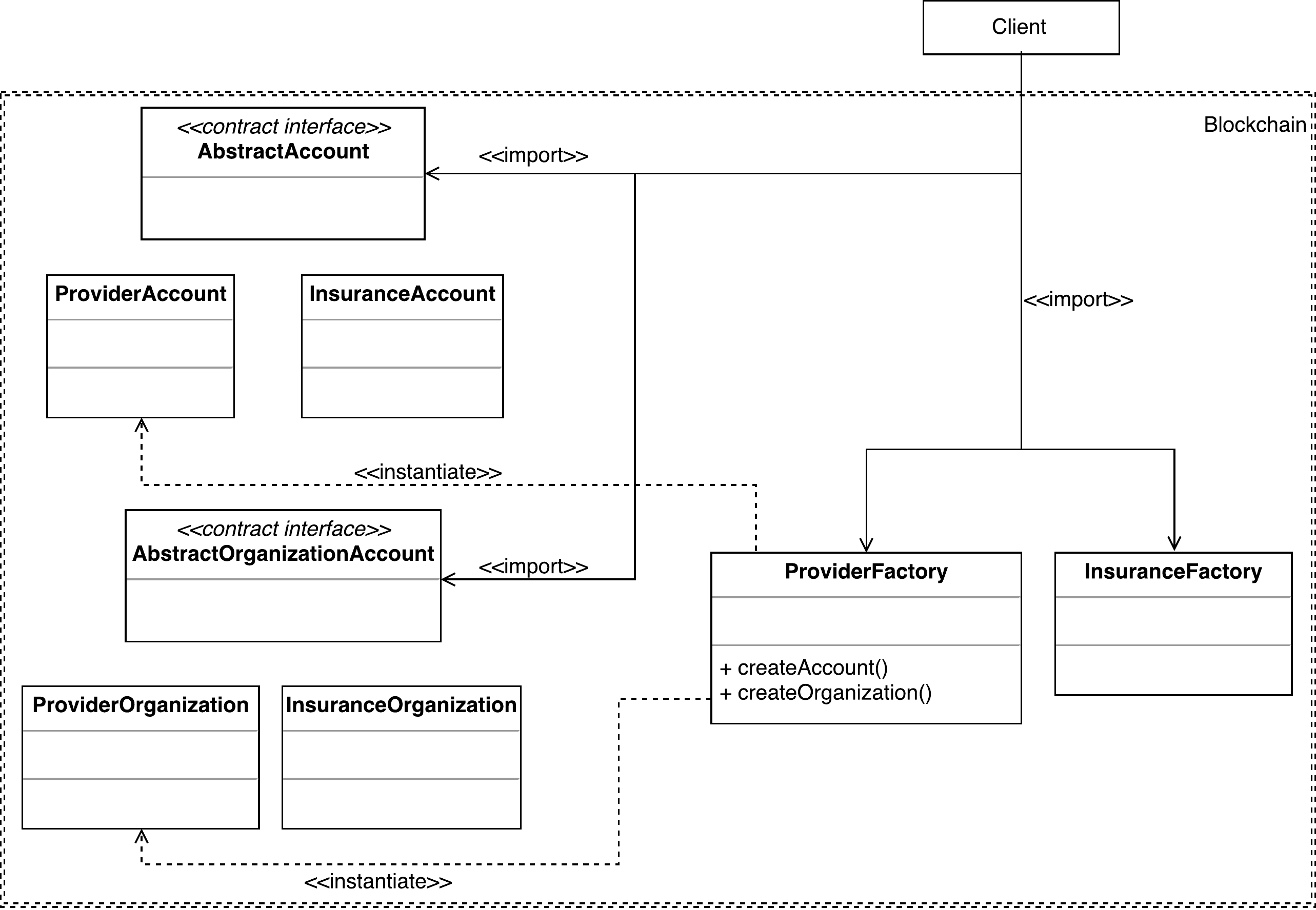}
\caption{Structure of an Application Without Applying the Abstract Factory Pattern}
\label{noabsfac}
\end{figure}

Without this factory contract, the client interested in creating department accounts for the entire organization has to access each of the specific account creation factories and make a lot of if-else decisions at runtime.  The tight coupling makes it cumbersome for the client since it needs to know all the implementation details in order to use each factory's methods correctly.  As shown in Figure~\ref{noabsfac}, to create an account for a care provider and insurance departments, the client imports both the ProviderFactory and the InsuranceFactory and invokes the createAccount() and createOrganization() methods for each account creation.  As an organization scales out and up with more departments and subdivisions, the client will be overwhelmed with these detailed implementations.  

\textbf{Intent.}  
\begin{itemize}
\item Defines a contract "interface" for creating a family of related or dependent contracts without having to explicitly specify their class types
\item Enables encapsulation through the hierarchical construction structure
\end{itemize}
 
\textbf{Applicability.}  The Abstract Factory pattern should be applied in the following scenarios:
\begin{itemize}
\item When the system needs to be independent from how different contracts in the system are created, such as creations of various multi-level user account contracts
\item When the system can be configured to work with families of related contracts, such as allowing interactions between a ProviderAccount from the AbstractAccount type and a ProviderOrganization from the AbstractOrganizationAccount type

\item When a library of contracts needs to be created that is relevant to the interface but not concrete contract implementations 
\end{itemize}

For example, in a healthcare blockchain app, a "super" contract Account could be used to define some common logic used by all types of user accounts (e.g. pharmacies, physicians' offices, insurance companies).  Some examples of such common logic may be patient data access, payment handling, and prescription lookup.  At creation time, the Account contract may not be able to anticipate all user contract types, so applying this pattern can allow future derived contracts to be created at runtime.  Specific and concrete implementations of the logic functions can be deferred to individual derived entity contracts.  
In the case of payment handling logic, for instance, a derived pharmacy contract could bill a pharmacy benefit management organization for a patient's prescriptions purchase; a physician's office derived contract may send bills associated with the patient visit to a medical insurance group; while an insurance group derived contract would then calculate the payment amount based on patients' benefit packages.  
 
\begin{figure}[bp]
\centering
\includegraphics[width=0.6\textwidth]{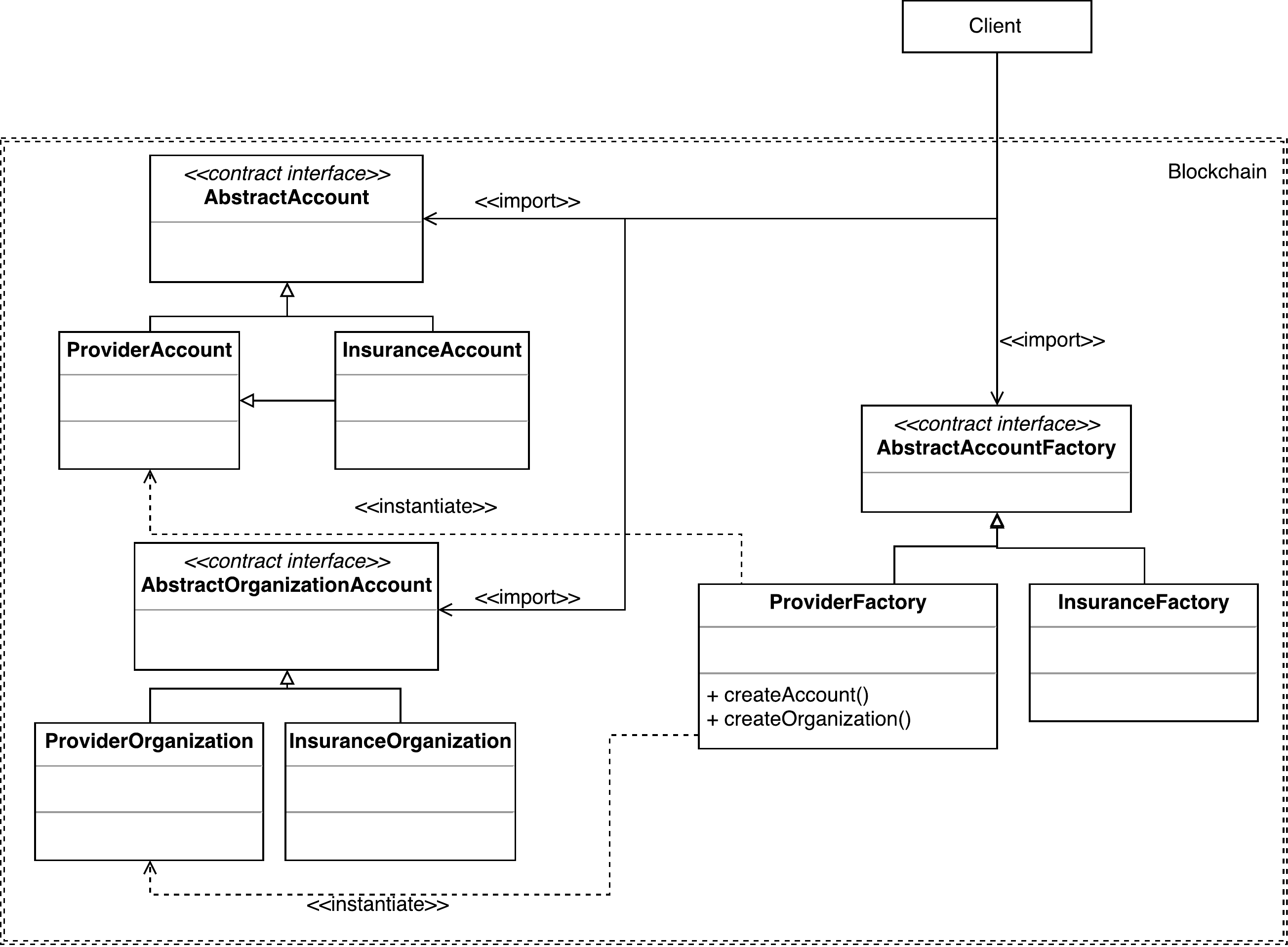}
\caption{Structure of an Application with the Abstract Factory Pattern}
\label{absfac}
\end{figure}

\textbf{Contract Structure.} As shown in Figure~\ref{absfac}, functions createAccount() and createOrganization() in the AbstractAccountFactory contract interface instantiate account contracts of a type created by the concrete implementation that handles the method call and returns the address to the newly created contract object. 

\textbf{Consequences.} The Abstract Factory pattern introduces a weak coupling between the application and concrete contract implementations, facilitating future contract extensions that share some commonality and also supporting customizations to the concrete contracts.  The downside of applying this pattern in the blockchain, however, is the extra cost incurred by an added layer of indirection in terms of storage for the interface contract and computation of instantiating the interface contract.  
 
\begin{figure}[tp]
\centering
\includegraphics[width=0.7\textwidth]{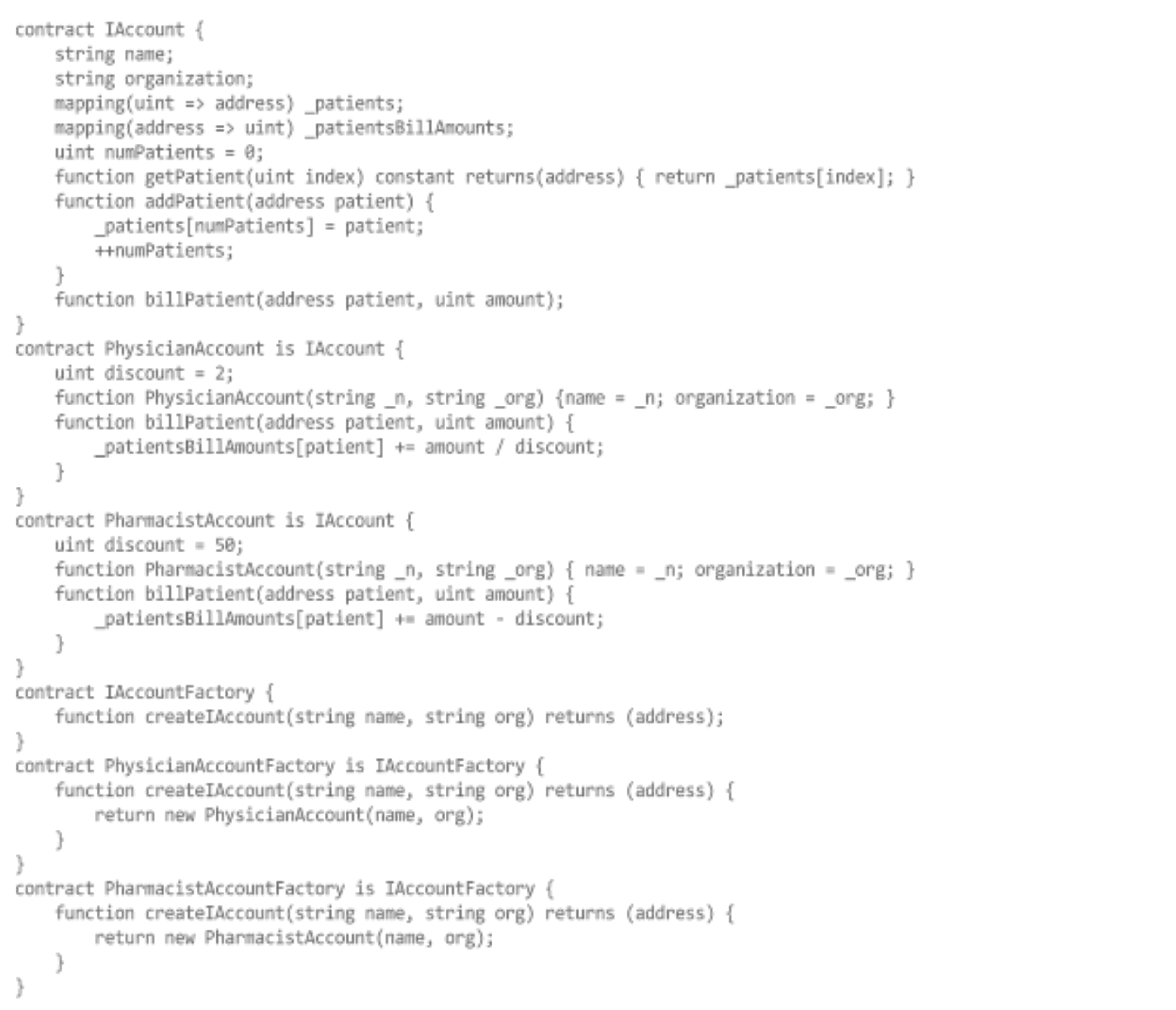}
\label{code1}
\end{figure}

\textbf{Sample Code.} The following code (Figure~\ref{code1}) from DASH provides an example of using the Abstract Factory pattern to create a phyisician account contract and a pharmacist account contract.

\subsubsection{Applying Flyweight to Address Interoperability Challenge 2}\mbox{}
 
\textbf{Motivation.} In order to achieve blockchain's transparency and immutability properties, all of the data and transaction records are maintained in the blockchain by replicating and distributing to every node in the network.  It is important to limit the amount of data stored in the blockchain to avoid high cost of data storage and unattended data when it is no longer needed.

For example, if a blockchain is being used to store patient billing data, there will be millions of patients stored in the blockchain. Moreover, each patient will require storage of their insurance information, which will include details, such as their ID\#, insurance contact information, coverage details, and other aspects that the provider needs to bill for services. Capturing this huge amount of information for every patient can generate a large amount of data in the blockchain.

The Flyweight pattern can aid in resolving the tension in this example between needing to store detailed insurance and billing information and the desire to minimize the information stored in the blockchain. Since most patients are covered by one of a relatively small subset of insurers (in comparison to the total number of patients), there is a substantial amount of intrinsic non-varying information that is common across patients, such as the policies on what procedures are covered by their insurance policy. Each insurance policy may cover 10,000s or 100,000s of patients and this information can potentially be reused and shared across patients. The details on what a particular policy covers are common across every patient with that policy. However, in order to bill for a service, this common intrinsic information needs to be combined with extrinsic information that is unique to each patient, such as the patient's ID\#.

Using the Flyweight pattern in smart contracts, intrinsic data shared by patients is stored in the common contract, while extrinsic data is stored in a separate contract representing that specific patient.  The contract with intrinsic data can also store references to specific patient contracts sharing the data.  When retrieving complete billing information, a common method call can be invoked to return the combined intrinsic and extrinsic data.  
 
\textbf{Intent.}
\begin{itemize}
\item Reuses existing similar contracts by storing them and create new contracts when no matching contract is found 
\item Supports a large number of contracts, such as account contracts, that have some common states where other parts vary
\end{itemize}
 
\textbf{Applicability.}
This pattern applies to a program that needs to handle a lot of contract objects that share some common states that can be externalized while other internal states vary.  It is also often used in combination with a factory that checks a pool of flyweights to determine if a flyweight object already exists in the pool.  In a healthcare system, each of many account types would contain a large number of detailed accounts that share some common metadata and/or accessing methods.  By applying the Flyweight pattern, shared representations can then be de-encapsulated in the Account "flyweight" factory only once, avoiding an exorbitant amount of memory usage from saving repeated data in all accounts. In addition, a flyweight factory can act as a Registry that maintains a mapping between unique user identifiers and the referencing addresses to the each account contract, preventing user account duplications.  At account creation, only if no account with the provided identifier exists in the registry does the factory create a new Account and return its address; otherwise the address associated with the existing Account contract will be returned.  A flyweight can have its own states and operations not shared with other objects. 
	A further substantial advantage of the flyweight pattern in the context of blockchains is that data is immutable once written. If insurance policy details are stored in each patient's contract directly, the cost to change a policy detail will be immense, since it will require rewriting a huge number of impacted contracts. The flyweight pattern also helps to minimize the cost of changes to the intrinsic state in blockchain applications.
 
\begin{figure}[bp]
\centering
\includegraphics[width=0.6\textwidth]{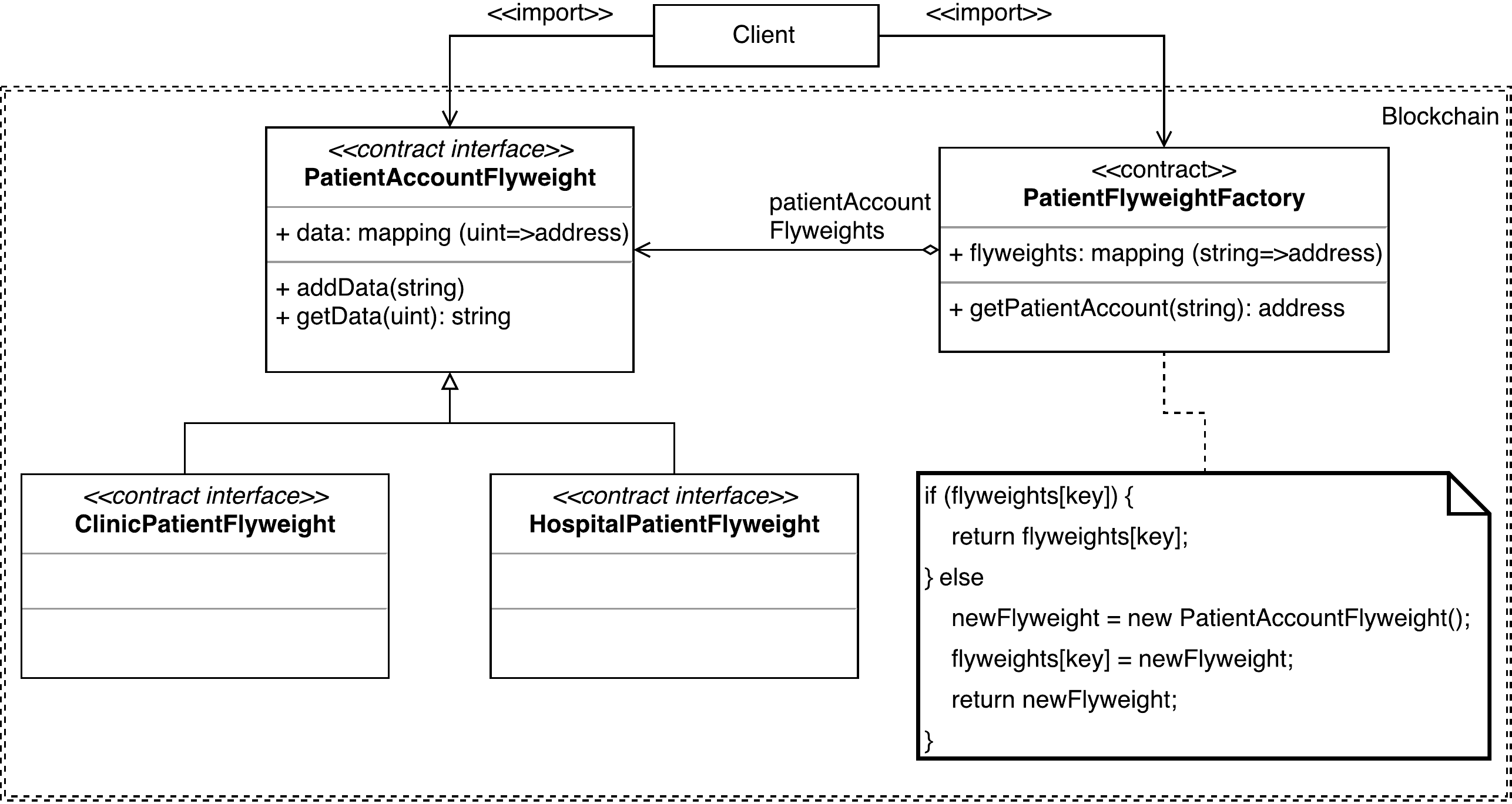}
\caption{Structure of Applying the Flyweight Pattern}
\label{flyweight}
\end{figure}

\textbf{Contract Structure.}
As shown in Figure~\ref{flyweight}, AbstractPatientAccount acts as a flyweight interface that defines intrinsic data structure and functions for concrete flyweights.  PatientAccountFactory, the flyweight factory keeps a records of created patient account contracts of all account types and creates a new flyweight when it does not already exist in the accounts mapping.

\textbf{Consequences.}
Applying the Flyweight pattern can provide better management to the large object pool, such as user accounts in the example above.  It minimizes redundancy in similar objects and, in the meantime, maximizes data and operation sharing.  Similar to the drawback of applying the Abstract Factory pattern, however, Flyweight incurs additional overhead to the client because of the extra layer of complexity.  Factory operation that returns the flyweight contract may also take longer to execute because it becomes another transaction to verify and included in the blockchain before a valid address can be available.  However, the computation delay may still be outweighed by the efficiency in management provided by Flyweight. 
 
\begin{figure}[tp]
\centering
\includegraphics[width=0.7\textwidth]{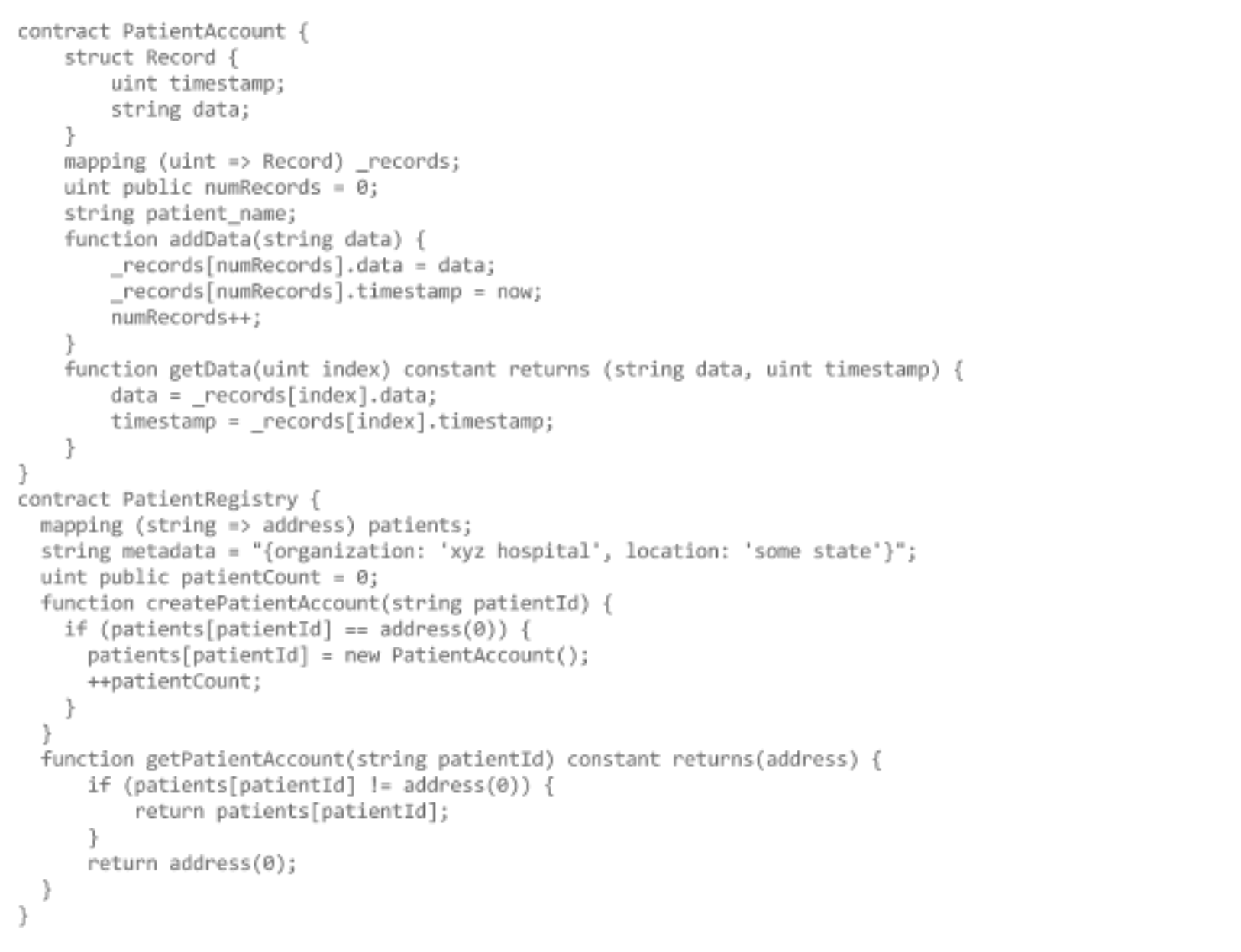}
\label{code2}
\end{figure}

\textbf{Sample Code.}
The following code (Figure~\ref{code2}) from DASH uses a Flyweight to maximize sharing of data between different patients and act as a registry to store each unique patient's account contract address.

\subsubsection{Applying Proxy to Address Interoperability Challenge 3}\mbox{}
 
\textbf{Motivation.} A fundamental aspect of a blockchain is that all data stored in the blockchain is public, immutable, and verifiable. For financial transactions that are focused on proving that transfer of an asset occurred, these properties are critical. However, when the goal is to store data in the blockchain, it is important to understand how these properties will impact the use case.
	
For example, storing patient data in a blockchain can be problematic, since it requires the data to be public and immutable. Although data can be encrypted before being stored in the blockchain, should all patient data be publicly distributed in the blockchain to all other parties for verification? Even if encryption is used, it is possible that the encryption technique may be broken in the future or that bugs in the implementation of the encryption algorithms or protocols used may lead to the information potentially being decryptable in the future. The immutability, however, prevents owners of the data from removing the data from the blockchain if a security flaw is ever found. Many other scenarios, ranging from discovery of medical mistakes in the data to changing standards may necessitate the need to change the data over time.

In scenarios where the data may need to be changed, the public and immutable nature of the blockchain creates a fundamental tension that needs to be resolved. On the one hand, the healthcare providers would like the data to be incorruptible so that it cannot be tampered with. At the same time, providers want the data changeable and private to protect patient privacy and account for possible errors. 

The Proxy pattern is a well-known software pattern that can be applied to blockchain-based data storage to resolve the tension created by the public and immutable aspects of the blockchain. Using the proxy pattern with a blockchain, a proxy contract is created to provide some lightweight representation or placeholder for the data with more intensive computation (such as acquiring data from off-blockchain storage via an Oracle~\cite{oraclize2017}).  The proxy contract can expose some simpler metadata of a patient and later refer to the heavyweight implementation on demand to obtain the real data object.  Each read request and modification operation of the data store can be logged in an audit trail that is transparent to the entire blockchain network for verification against data corruption.  When the proxified contract (with heavyweight implementation) is updated with new storage option, for instance, interface to the proxy contract can remain the same, encapsulating detailed implementation variations.    
 
\textbf{Intent.}
\begin{itemize}
\item Provides a surrogate or placeholder contract for a health data storage object for another object to control access to it
\item Supports distributed or controlled access to sensitive information using an additional level of indirection
\item Creates a wrapper to protect data storage from undue complexity
\end{itemize}

\textbf{Applicability.}
Remote proxies can be useful in an Ethereum healthcare application as they provide a local representation for an object that is in a different address space.  In the storage case, patient medical data could be stored either inside contracts on the blockchain as contract states or event logs or externally off the blockchain as encrypted file objects.  In this scenario, a remote proxy can be applied on the application server side to create a reference or wrapper to the actual data storage and later retrieve the data when needed.  Instead of directly applying a remote proxy inside a smart contract in Solidity, it would be applied in the healthcare application's web server to create a wrapper for the contract object containing relevant data if it is used as storage.

A protective proxy can also be used to control access to the original sensitive object.  For instance, to prevent unauthorized users on the blockchain to change the states of a patient data object, it can set up a proxy to do some permission checking prior to forwarding the change request.
 
\begin{figure}[bp]
\centering
\includegraphics[width=0.6\textwidth]{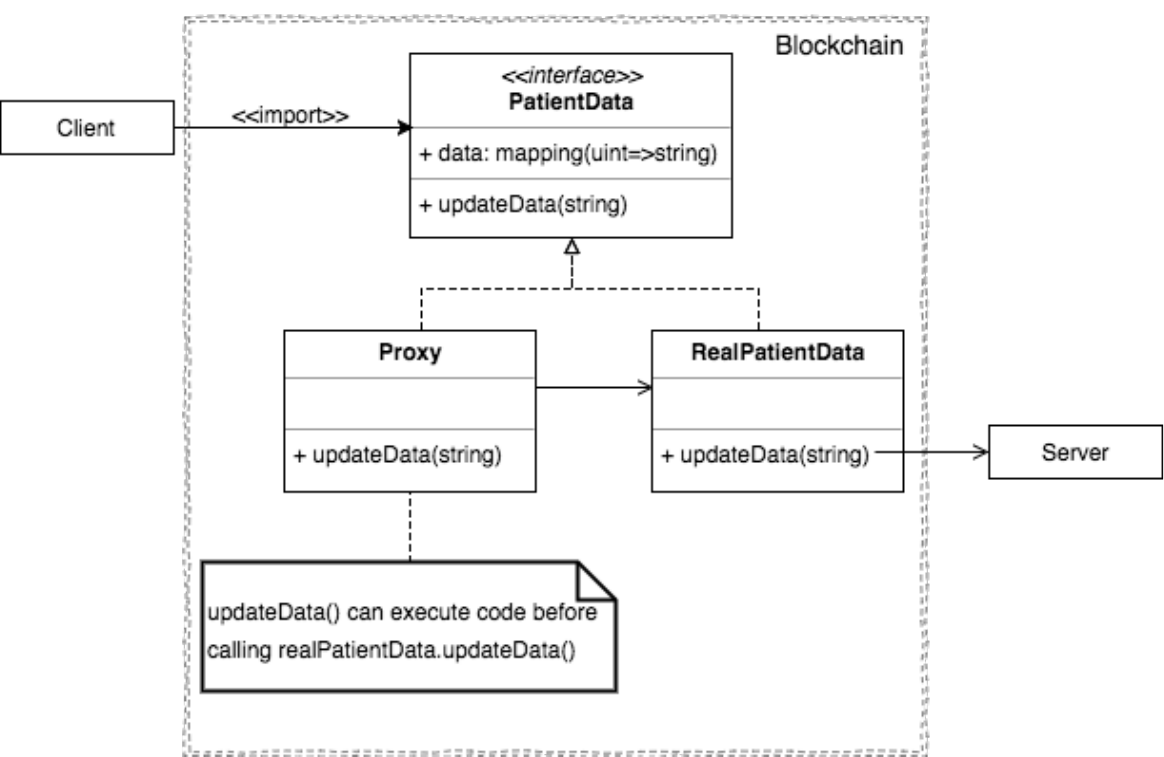}
\caption{Structure of Applying the Proxy Pattern}
\label{proxy}
\end{figure}

\textbf{Contract Structure.}
As shown in Figure~\ref{proxy}, Proxy and RealPatientData implement the same interface, with Proxy being the lightweight surrogate to RealPatientData, the heavyweight implementation.

\textbf{Consequences.}
A proxy object can perform simple housekeeping or auditing tasks by storing some commonly used metadata in its internal states without having to perform heavy operations such as decrypting a file.  It follows the same interface as the real object and can execute the original heavyweight function implementations on demand.  It can also hide information about the real object from the client to protect patient data privacy.  Since Proxy introduces another layer of indirection, however, it could cause disparate behavior when the real object is accessed directly by some client while the surrogate is accessed by others.  In addition, contracts in Solidity are instantiated by other contracts or transactions, meaning that the proxy can only reference the real contract with its address on the blockchain, which defeats the purpose of using a protection proxy to hide the real object from the client.  Nevertheless, both remote and protective proxies can be used on the server side of the application that accesses these contracts. 
 
\begin{figure}[tp]
\centering
\includegraphics[width=0.7\textwidth]{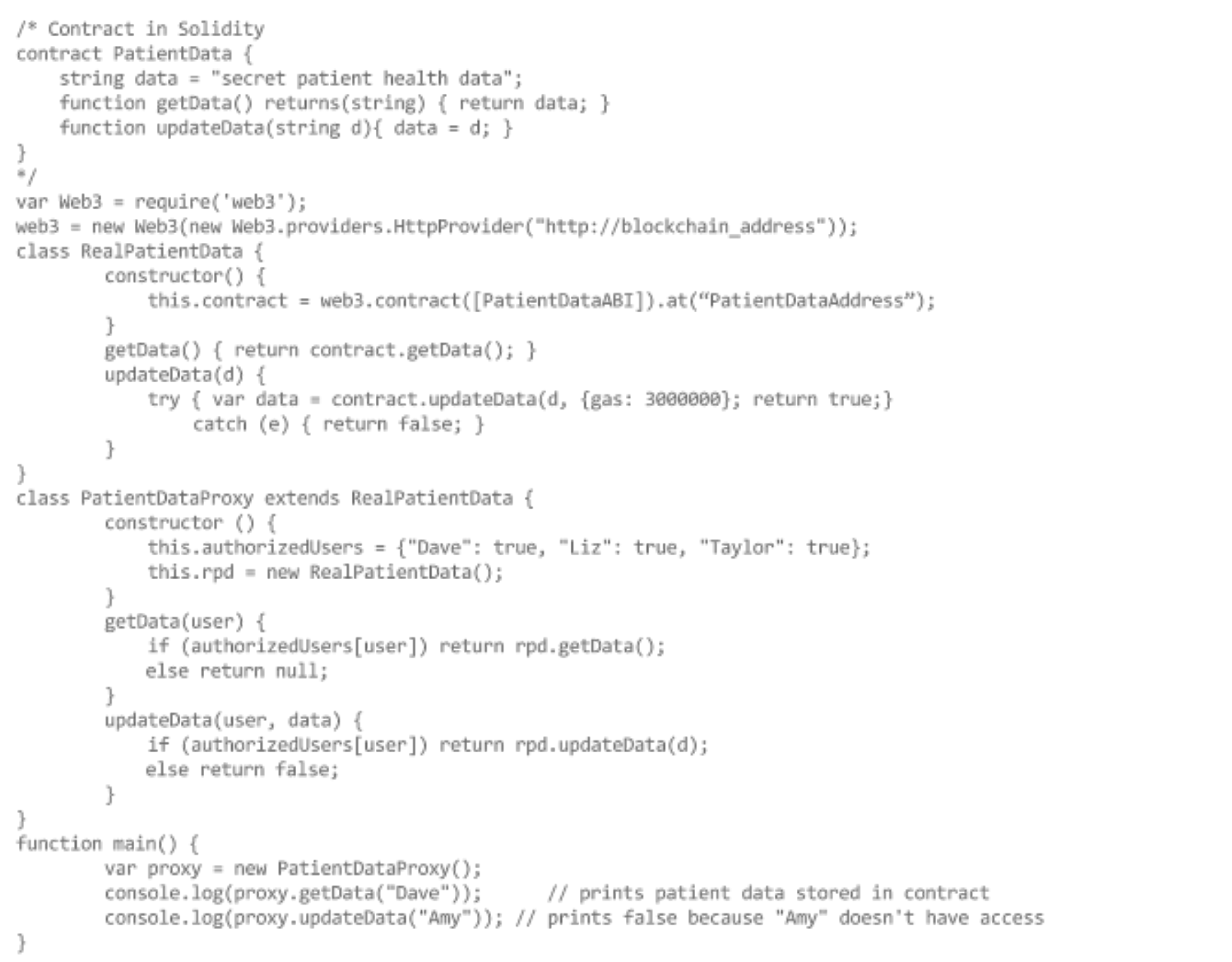}
\label{code3}
\end{figure}

\textbf{Sample Code.}
The following code (Figure~\ref{code3}) from our DASH app uses a protective Proxy to control access to the smart contract implementation of data acquisition.  

\subsubsection{Applying Publisher-Subscriber to Address Interoperability Challenge 4}\mbox{}
 
\textbf{Motivation.} 
	The Ethereum blockchain maintains a public record of contract creations and operation executions along with regular cryptocurrency transactions.  The availability of information makes blockchain a more autonomous approach to improve the coordination of patient care across different teams (e.g. physicians, pharmacists, insurance agents, etc) who would normally communicate through various channels with a lot of manual effort, such as through telephoning or faxing.  Although, from a continually growing list of records, directly capturing any specific health-related topic of occurred events would require a lot of transaction receipt lookups and topic filtering, which requires non-trivial computation and may result in delayed responses.  

For instance, when a blockchain application is used to support coordinated care through patient self reporting of illness or prescription request, it is important to relay the report and any follow up procedure to and from the associated care provider offices in a timely manner.  In this case, the Publisher-Subscriber pattern can assist in broadcasting the information only to care providers that subscribe to events relating to this patient.  It solves the issue of constant information filtering by actively monitoring patient activities and sending notifications to the patient's care team as the activities take place.  To avoid computation overhead on the blockchain, the actual processing of patient activities data can be done off-chain by a back-end server.  When receiving the events of interest, the subscribers can then pass the heavy computation tasks to the server.

\textbf{Intent.}
\begin{itemize}
\item Creates a messaging infrastructure using separation of concerns by allowing publishers to create messages and subscribers to receive messages of their interests.

\item Enables interoperability within a healthcare ecosystem so that participating departments can be notified of health-related events they are concerned with.

\end{itemize}

\textbf{Applicability.}
This pattern should be applied when a state change in one object must be reflected in another object without keeping the objects tightly coupled and when the system is to be enhanced with new topics or subscribers with minimal changes to the implementation.

In a health blockchain application, a care provider may prescribe a patient a list of medications during a visit, sending (publishing) a message of the prescription update.  Involved parties, the patient's pharmacy and insurance company may subscribe to messages related to this patient's prescription activities in order to receive this update from the provider's office.  Other parties that do not have direct associations with this patient, such as a hospital in a different geographic area, may choose not to subscribe to this patient's activities to avoid the overhead of receiving irrelevant information.  

The Publisher-Subscriber pattern can be realized in a smart contract because the publisher itself can handle event notifications to subscribed parties, but it is more cost-efficient to be implemented by the health application server. 
 
\begin{figure}[bp]
\centering
\includegraphics[width=0.6\textwidth]{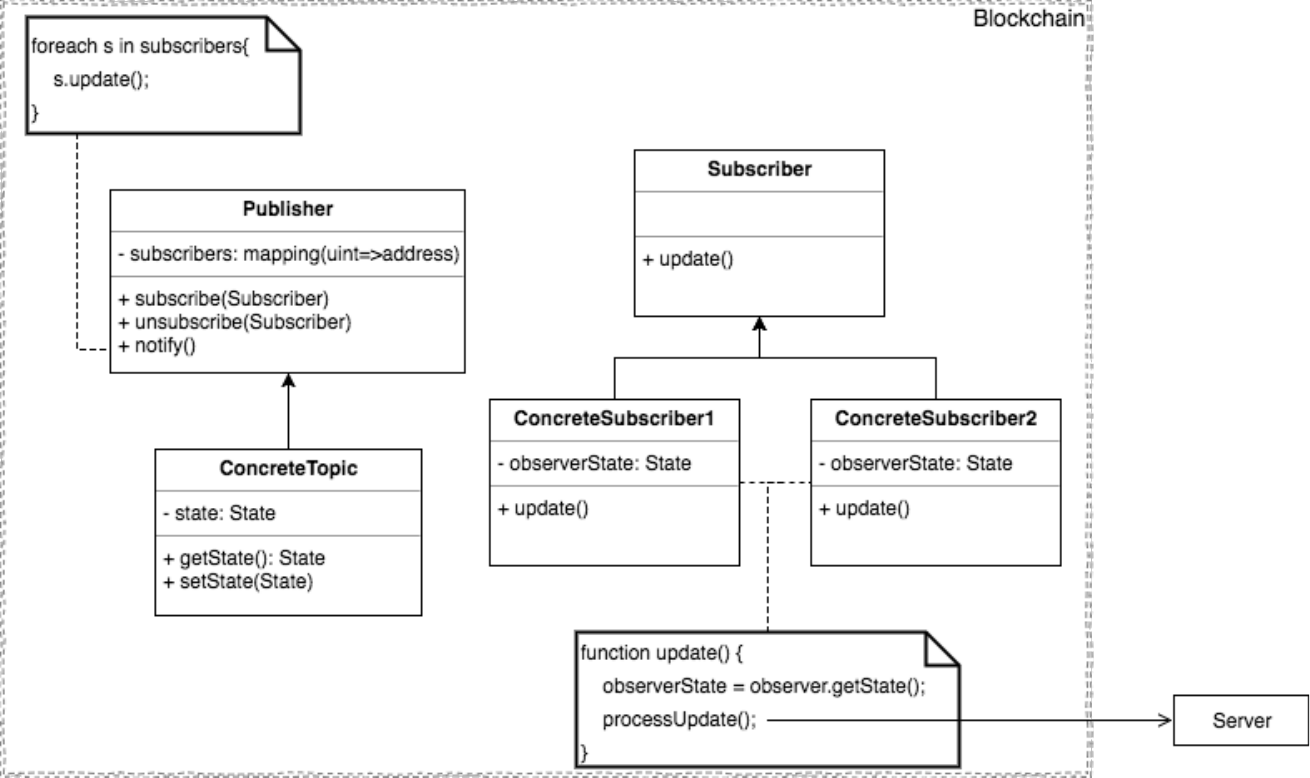}
\caption{Structure of Applying the Publisher Subscriber Pattern}
\label{pubsub}
\end{figure}

\textbf{Contract Structure.}
As shown in Figure~\ref{pubsub}, publisher keeps track of a list of subscribers. When an event occurs, publisher notifies all the subscribers of the event, and subscribers can then process the event update..

\textbf{Consequences.}
The limitations of realizing the Publisher-Subscriber pattern on the blockchain include (1) delays in updates received by subscribers due to the extra step of validation required by the blockchain infrastructure, (2) high computational costs associated with filtering very fine-grained topic subscriptions on-chain, and (3) high executional costs associated with notifying events.  With a much faster mining process on the Ethereum blockchain compared to Bitcoin, each block can be mined and added to the blockchain roughly every 12 seconds in Ethereum.  The order of which transactions are to be executed is determined by the miners based on the transaction fees paid by the senders and how long transactions have been in the transaction pool.  Transactions with the highest priority will be executed first and added to the blockchain sooner, which could cause delays in publishers sending out messages and subscribers receiving messages of their interest.  Every computation occurred on the blockchain will be charged some amount of "gas", so the more intensive the computations are, the more costs to realize them.  If subscribers want to receive very fine-grained message topics, the amount of computation for filtering out the messages sent out by publishers may cost an exorbitant amount, resulting in either failure to publish messages due to gas shortage or unacceptable implementation costs.  A workaround could be to have broader topics with fewer filter requirements on-chain and handle more detailed message filtering off Blockchain.  
 
\begin{figure}[tp]
\centering
\includegraphics[width=0.7\textwidth]{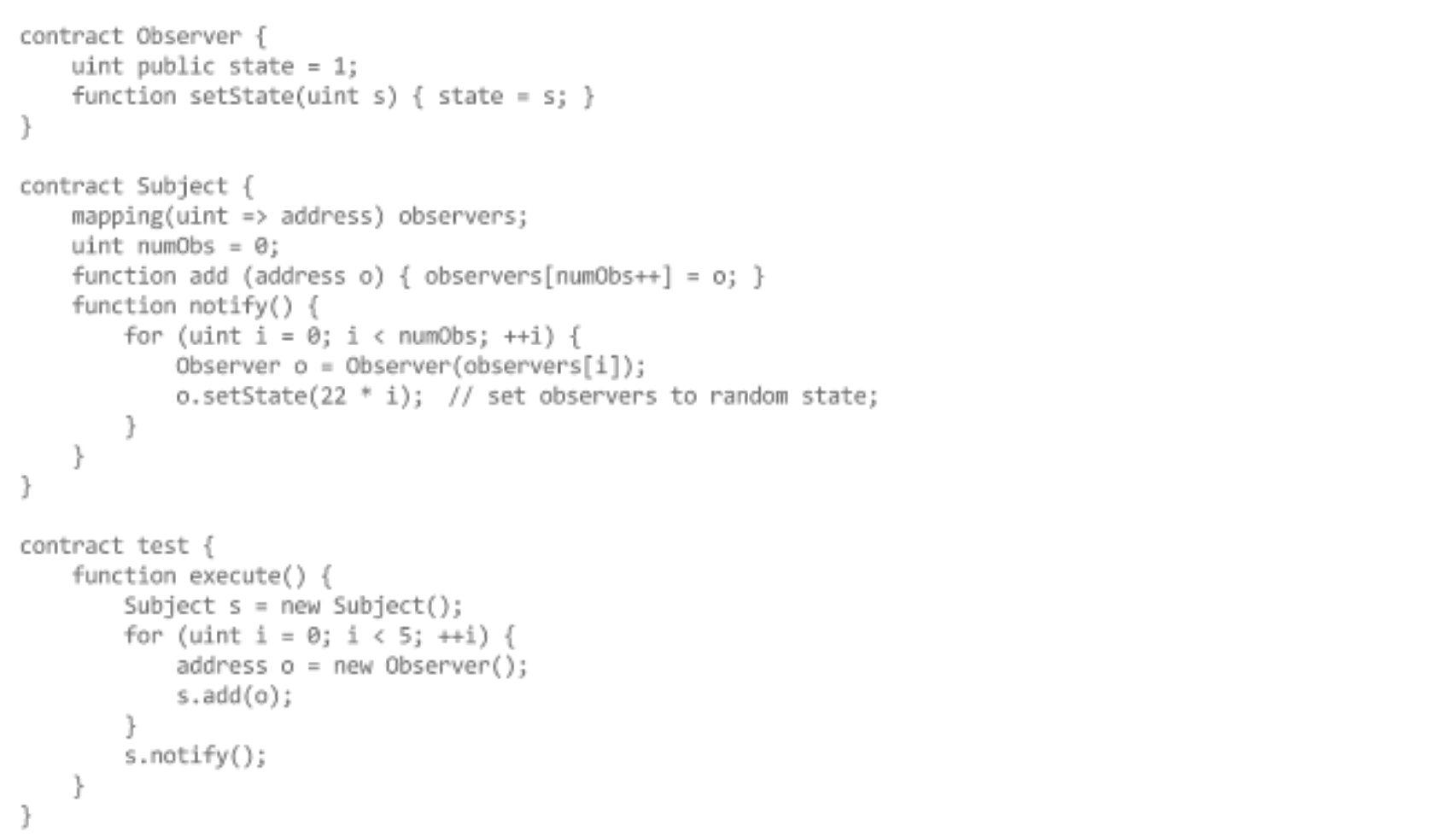}
\label{code4}
\end{figure}

\textbf{Sample Code.}
In the following example from our DASH app (Figure~\ref{code4}), 5 observers are subscribed to the Subject contract.  After the execute method is invoked, each subscriber will receive an update that is reflected by the change of value in the variable "state". 

Despite the drawbacks of added complexity and computation costs, it is considered good practice to modularize contract code and server-side code that accesses contracts with appropriate software patterns to allow easier extensibility and flexible customizations to each major component in a healthcare blockchain application.  Software patterns also provide separation of concerns through loose coupling between components, allowing components to access shared methods and artifacts while still encapsulating variabilities in detailed implementations.  An interoperable healthcare chain should enable seamless communications and interactions between components, which can be made robust and customizable with careful evaluations of overall workflow and applicable software patterns. 

On the application's server end, we applied the Publisher-Subscriber pattern to allow health providers to subscribe to patient account activities.  For instance, when a patient requests a prescription drug through the web portal, all the subscribed providers of that patient will be notified of the event, but only permissioned providers will be able to respond to the prescription request via an update to the patient data.   

When writing data to a patient's account, a permissioned care provider first queries the patient registry for the location of the patient account contract.  If the patient does not yet exist in the registry mapping, a new Patient Account contract will be automatically created and registered in Patient Registry with the patient ID, using the Flyweight and Abstract Factory patterns.  The registry then returns the address pointing to the specified patient account.  With the contract address, the provider will then be able able to write new data to the patient health records by adding the new data entry in the contract internal data array.  Similarly, when reading a patient record, a provider must first query Patient Registry with a patient ID.  If the requested patient exists, Patient Registry returns the address of Patient Account contract from which patient data can be retrieved; otherwise, it creates a contract account for the patient and returns the new address where the data field is empty.  

For testing and backup purposes, we have also created a separate patient record database in MongoDB.  To decouple our data store choices from other operations, on the application's server end, we applied the Proxy pattern to create a wrapper for a generic database object and refer to the Proxy until we need to obtain the actual data.

\subsection{Related Work}
Although relatively few papers focus on realizing software patterns in blockchains, many papers relate to healthcare blockchain solutions and applying software design principles in this space.  This section gives an overview of related research on (1) the challenges of applying blockchain-based technology in the healthcare space and innovative implementations of blockchain-based healthcare systems and (2) design principles and recommended practice for blockchain application implementations.

\subsection{Challenges of healthcare blockchain and proposed solutions.}
Ekblaw et. al proposed MedRec as an innovative, working healthcare blockchain implementation for handling EHRs, based on principles of existing blockchains and Ethereum's smart contracts~\cite{ekblaw2016case}.  The MedRec system uses database "Gatekeepers" for accessing a node's local database governed by permissions stored on the MedRec blockchain.  Peterson et. al presented a healthcare blockchain with a single centralized source of trust for sharing patient data, introducing "Proof of Interoperability" based on conformance to the Fast Healthcare Interoperability Resources (FHIR) protocol as a means to ensure network consensus~\cite{peterson2016blockchain}.

\subsection{Applying software design practice to blockchain.} Porru et. al highlighted evident challenges in state-of-the-art blockchain-oriented software development by analyzing open-source software repositories and addressed future directions for developing blockchain-based software, focusing on macro-level design principles such as improving collaboration, integrating effective testing, and evaluations of adopting the most appropriate software architecture~\cite{porru2017blockchain}.  Bartoletti et. al surveyed the usage of smart contracts and identified nine common software patterns shared by the studied contracts, e.g. using "Oracles" to interface between contracts and external services and creating "Polls" to vote on some question.  These patterns summarize the most frequent solutions to handle some repeated scenarios~\cite{bartoletti2017empirical}. 
\section{Concluding Remarks}
 	This paper provided an overview of the blockchain platform and described the motivations for applying blockchain technology to solve healthcare interoperability issues, focusing on (1) maintaining evolvability while minimizing integration complexity, (2) minimizing data storage requirements, (3) balancing integration ease with security concerns, and (4) tracking relevant health changes across large patient populations.  We presented a case study of our DApp for Smart Health (DASH) to highlight the implementation challenges that rose when we attempted to extended the app, such as tightly coupled designs, duplicated resources, and lack of scalability.  To address interoperability challenges while providing solutions to the implementation challenges, we identified and applied four software patterns to DASH to improve its design and functionality.  
    
	From this study we learned that the public, immutable, and verifiable properties of the blockchain allow for a more interoperable environment that is not easily achieved using traditional approaches that mostly rely on a centralized server or data storage.  Important design decisions need to be made in advance to better take advantage of the smart contract support while avoiding much computation and storage overhead.  Combining good software design practice with these unique properties of the blockchain can create apps that are more modular and easier maintenance of the smart contracts.  In particular, we applied Abstract Factory, Flyweight, Proxy, and the Publisher-Subscriber patterns to DASH to decouple the creation and access of entities, maximize sharing of resources, and improve application scalability.
    
	Our future work will extend the app described in Sections 3 and 4 to delve into the challenges and pinpoint the most practical design process in designing a healthcare blockchain architecture.  In addition, we will explore the potential application of other software patterns to handle various challenges, such as security, privacy, dependability, and performance. One approach is to conduct experiments to evaluate the efficacy of applying software patterns to this architecture compared to other alternative designs.  We will also investigate extensions of the blockchain from a healthcare perspective, such as creating an alternative health chain that exclusively serves the healthcare sector.

% Head 1

% Bibliography
\bibliographystyle{ACM-Reference-Format-Journals}
\bibliography{acmlarge-sam}

% History dates
% {Initial submission}{Conference Version}{Submission after having considered Writers' Workshop feedback}
% \received{May 2015}{September 2015}{February 2015}

% \elecappendix

\end{document}